# Tomography of evolved star atmospheres

**Kateryna Kravchenko**

Max Planck Institute for extraterrestrial Physics, Giessenbachstraße 1, 85748, Garching, Germany

Cool giant (AGB) and supergiant (RSG) stars are among the largest and most luminous stars in the Universe. They were extensively studied during last few decades, however their relevant properties like photometric variability and mass loss are still poorly constrained. Understanding these properties is crucial in the context of a broad range of astrophysical questions including chemical enrichment of the Universe, supernova progenitors, and the extragalactic distance scale.

Atmospheres of AGBs and RSGs are characterized by complex dynamics due to different interacting processes, such as convection, pulsation, formation of molecules and dust, and the development of mass loss. The state-of-the-art 3D radiative-hydrodynamics (RHD) simulations of AGBs and RSGs [1] are able to simulate some of those processes and produce a good agreement with the observed spectral features. However, the models lack constraints and need to be confronted to observables. A recently established tomographic method is an ideal technique for this purpose.

masks) containing lines forming at given, pre-specified ranges of optical depths. In order to correctly assess the depth of formation of spectral lines, the computation of a contribution-function was implemented by [3] in 1D radiative transfer code TURBOSPECTRUM [4], which performs spectrum synthesis from 1D static model atmospheres. The cross-correlation of the masks with observed or synthetic stellar spectra provides the velocity at different atmospheric depths. The tomographic technique was fully validated by cross-correlating a synthetic spectrum computed from a snapshot of a 3D RHD simulation [3] with a set of tomographic masks. The distribution of the line-of-sight velocity field throughout the atmosphere (known from the input model atmosphere) was nicely recovered (Figure 1). The following sections describe some of our applications of the tomographic method.

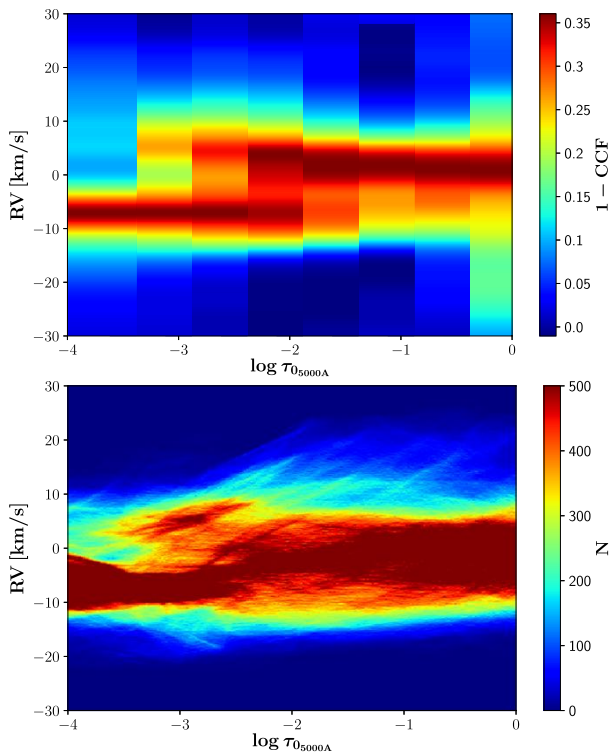

**Figure 1:** Reconstruction of velocity field as function of optical depth in the atmosphere. Top: cross-correlation of a 3D snapshot spectrum with the tomographic masks. Bottom: 3D snapshot velocity distribution.

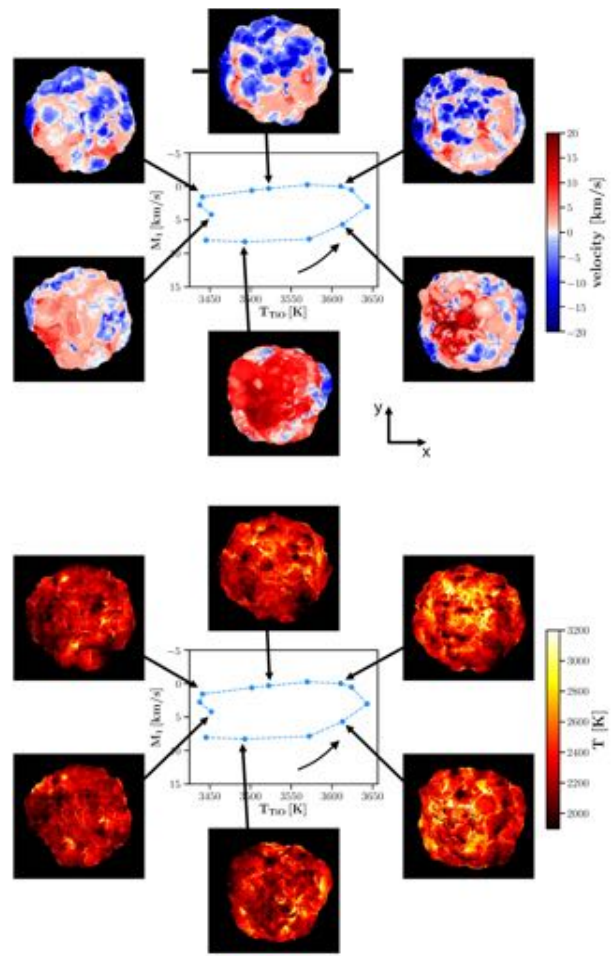

**Figure 2:** Evolution of velocity (top) and temperature (bottom) of the 3D simulation during a single photometric cycle represented by the hysteresis loop (center of each panel).

### Tomographic method

The tomographic method [2, 3] is based on cross-sectioning the stellar atmosphere in order to reconstruct the velocity field for each atmospheric layer. The method relies on the design of spectral templates (or

### Interpretation of photometric variability in RSG stars

RSG stars are characterized by irregular photometric variations on timescales of hundreds of days. This ob-served variability can correspond to a broad range of processes: atmospheric pulsations and/or convective motions, magnetic activity, binarity. In order to identify the responsible mechanism, the tomographic method was applied to time-series of high-resolution (HR) optical spectra of two RSG stars: $\mu$ Cep [5] and Betelgeuse [6]. A phase shift was detected between the radial velocity (RV) and photometric (as well as effective temperature) variations. This phase shift results in hysteresis loops in the temperature-velocity plane with timescales similar to the photometric ones. The application of the tomographic method to snapshots from 3D RHD simulations of a RSG atmosphere revealed similar hysteresis-like behavior in the temperature-velocity plane (Figure 2) with the lifetime consistent with the photometric timescales. The temporal variation of simulated velocity and temperature structures along a hysteresis loop (Figure 2) is indicative of atmospheric convective motions. The appearance of bright warm regions on the surface is followed by the rising material at same locations, which accounts for the phase lag between velocity and photometric variations and, thus, hysteresis loops. Therefore, convection plays a major role in the photometric variability of RSG stars.

### Link between optical and geometrical depth scales

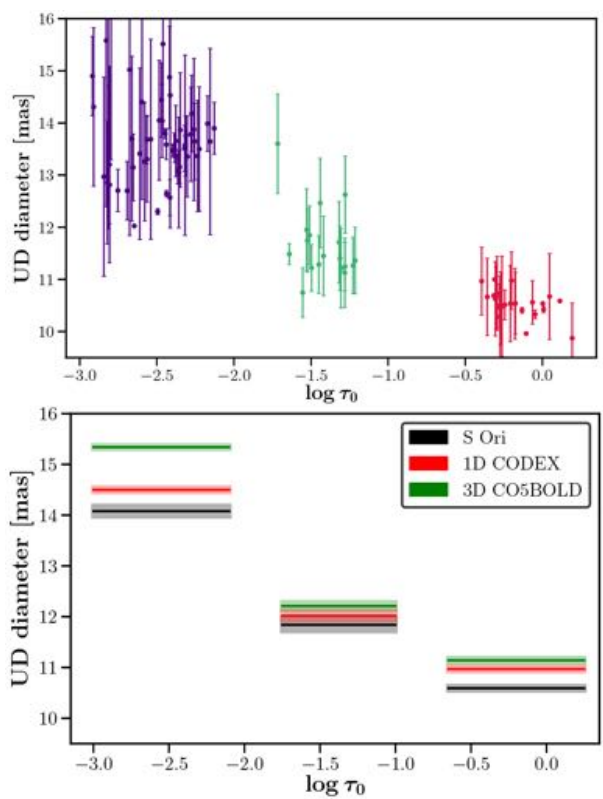

**Figure 3:** Relation between optical (x-axis) and geometrical (y-axis) depth scales for S Ori. Top: uniform disk (UD) fit of interferometric visibilities for each line of the three tomographic masks (different colors). Bottom: comparison with 1D and 3D dynamical models.

Current stellar atmosphere models of AGB and RSG stars provide a relation between optical and geometrical depth scales. However, this relation is sensitive to the stellar mass, which is ill-defined since these stars are undergoing a substantial mass loss. In addition, the models experience several limitations. For example, 3D RHD simulations are able to represent convection and pulsation processes, but they do not yet include wind, radiation pressure, and magnetic field, all of which may affect the structure and dynamics of the atmosphere. Thus, the knowledge of the relation between optical and geometrical depth scales from observations would open a way to test and constrain the models.

The observed relation between optical and geometrical depth scales can be derived by applying the tomographic method to HR spectro-interferometric (e.g. VLTI/AMBER or VLTI/GRAVITY) observations. By extracting visibilities (i.e. the contrast of interferometric fringes) at wavelengths contributing to the tomographic masks, we can measure the extension of different atmospheric layers [7]. Figure 3 illustrates our pilot study of the Mira-type AGB star S Ori [7] where the link between optical (provided by the tomographic masks) and geometrical (provided by fitting the interferometric visibilities) depth scales is nicely recovered and compared to dynamical model atmospheres.

### Conclusions

The tomographic method is a powerful and precise technique to probe different depths in stellar atmospheres. So far, its application to AGB and RSG stars allowed to interpret their photometric variability and recover a quantitative relation between optical and geometrical depth scales. The future prospects of the tomographic method include its application to a broader range of astrophysical objects, such as Cepheid variables and exoplanets, to reveal detailed structure and dynamics of their atmospheres.

### Short CV

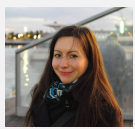

| | |
|---|---|
| 2013–2014: | Master in Astrophysics, Kharkiv, Ukraine |
| 2015–2019: | PhD in Science, Brussels, Belgium |
| 2019–2020: | ESO Fellow, Santiago, Chile |
| 2020–present: | Post-doctoral researcher, MPE, Garching, Germany |